\documentclass{article}

\usepackage{PRIMEarxiv}
\usepackage{amsmath} 
\usepackage[utf8]{inputenc} 
\usepackage[T1]{fontenc}    
\usepackage{hyperref}       
\usepackage{url}            
\usepackage{booktabs}       
\usepackage{amsfonts}       
\usepackage{nicefrac}       
\usepackage{microtype}      
\usepackage{lipsum}
\usepackage{fancyhdr}       
\usepackage{graphicx}       
\graphicspath{{media/}}     

\title{SocialRec: User Activity Based Post Weighted Dynamic Personalized Post Recommendation System in Social Media
}

\author{
  Ismail Hossain, Sai Puppala, Md Jahangir Alam, Sajedul Talukder \\
  Computer Science\\
  University of Texas at El Paso, TX, USA, 79902\\
  School of Computing \\
  Southern Illinois University Carbondale, IL, USA, 62901\\
  \texttt{\{ihossain, malam10\}@miners.utep.edu, sai.puppala@siu.edu, stalukder@utep.edu} \\
}

\begin{document}

\title{EVOLVE-X: Embedding Fusion and Language Prompting for User Evolution Forecasting on Social Media}

\maketitle


\begin{abstract}
 Social media platforms serve as a significant medium for sharing personal emotions, daily activities, and various life events, ensuring individuals stay informed about the latest developments. From the initiation of an account, users progressively expand their circle of friends or followers, engaging actively by posting, commenting, and sharing content. Over time, user behavior on these platforms evolves, influenced by demographic attributes and the networks they form. In this study, we present a novel approach that leverages open-source models Llama-3-Instruct, Mistral-7B-Instruct, Gemma-7B-IT through prompt engineering, combined with GPT-2, BERT, and RoBERTa using a joint embedding technique, to analyze and predict the evolution of user behavior on social media over their lifetime. Our experiments demonstrate the potential of these models to forecast future stages of a user's social evolution, including network changes, future connections, and shifts in user activities. Experimental results highlight the effectiveness of our approach, with GPT-2 achieving the lowest perplexity (8.21) in a Cross-modal configuration, outperforming RoBERTa (9.11) and BERT, and underscoring the importance of leveraging Cross-modal configurations for superior performance. This approach addresses critical challenges in social media, such as friend recommendations and activity predictions, offering insights into the trajectory of user behavior. By anticipating future interactions and activities, this research aims to provide early warnings about potential negative outcomes, enabling users to make informed decisions and mitigate risks in the long term.
\end{abstract}

\keywords{GPT-2, BERT, RoBERTa, User Evolution, Social Media, Join Embedding, Prompting, LLM}

\section{Introduction}
\begin{figure*}[h]
    \centering
    \includegraphics[width=0.7\textwidth]{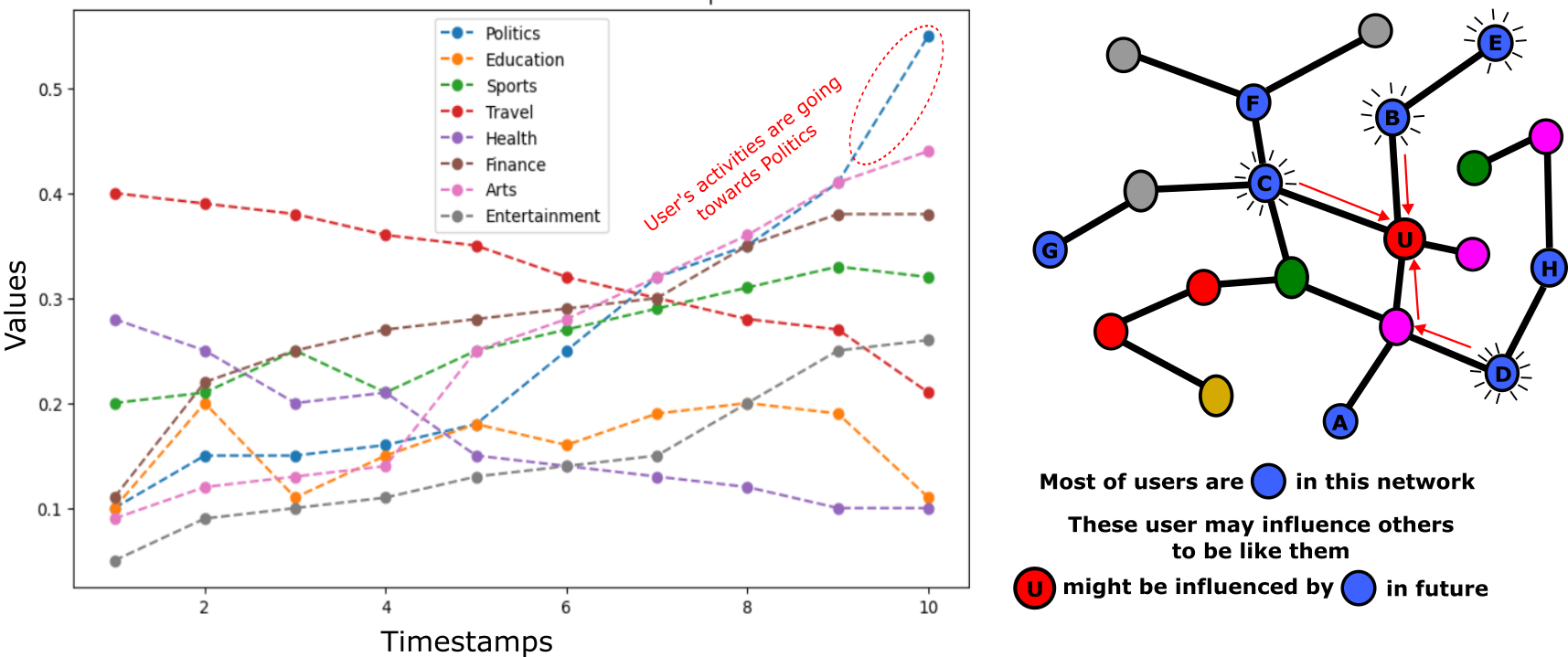}
    \caption{User habits on social media might change due to others who are in the same network}
    \label{fig:evolution-impact}
\end{figure*}

In the dynamic and ever-evolving landscape of social media, understanding the changes in user behavior is essential. User evolution encompasses alterations in interaction patterns, preferences, and social ties over time, driven by demographic attributes and the networks users form. The complexity of social media platforms, powered by sophisticated algorithms, user-generated content, and intricate networks, underscores the importance of analyzing these changes. To capture shifts in user behavior or activity on social media, we mapped the activity of a Facebook user over a decade. Figure~\ref{fig:evolution-impact} highlights transitions in user activities, such as increased engagement with political posts and a reduction in travel-related content. Another figure illustrates how user $U$'s behavior evolves through interactions with users $D, C, B, E$, demonstrating the role of social connections in driving behavioral evolution.

Researchers have extensively studied various aspects of online social media, uncovering insights into user engagement patterns, network dynamics, and the influence of algorithms on information dissemination~\cite{pariser2011filter}. The evolution within social networks includes diverse dimensions such as community evolution tracking~\cite{tracking_dakiche_2019}, changes in social power and lifecycle events~\cite{predicting_sharmeen_2015}, the impact of triadic closure on link formation~\cite{role_weng_2013}, and the temporal evolution of location-based social networks~\cite{evolution_allamanis_2012}. Cultural diversity~\cite{evolution_grabowski_2006} and shifts in roles or social status within networks~\cite{analyzing_lin_2009} further highlight the multifaceted nature of user evolution. Building on these foundations, our methodology integrates user network data, demographic information, user history, and engagement metrics to analyze the phases of user evolution in online social networks. By employing state-of-the-art techniques, we aim to uncover intricate patterns and emergent phenomena that drive behavioral changes. Central to this research is the critical question: \textit{How do users evolve within the complex web of platforms such as Facebook, Twitter, and Reddit, and what factors propel this evolution?} Our study seeks to decipher the mechanisms that govern user adaptation, examining how individual behaviors, network structures, and platform dynamics interconnect to shape user trajectories over time. These insights aim to inform the development of more adaptive and user-centric social media platforms.

\begin{figure*}[h]
    \centering
    \includegraphics[width=\textwidth]{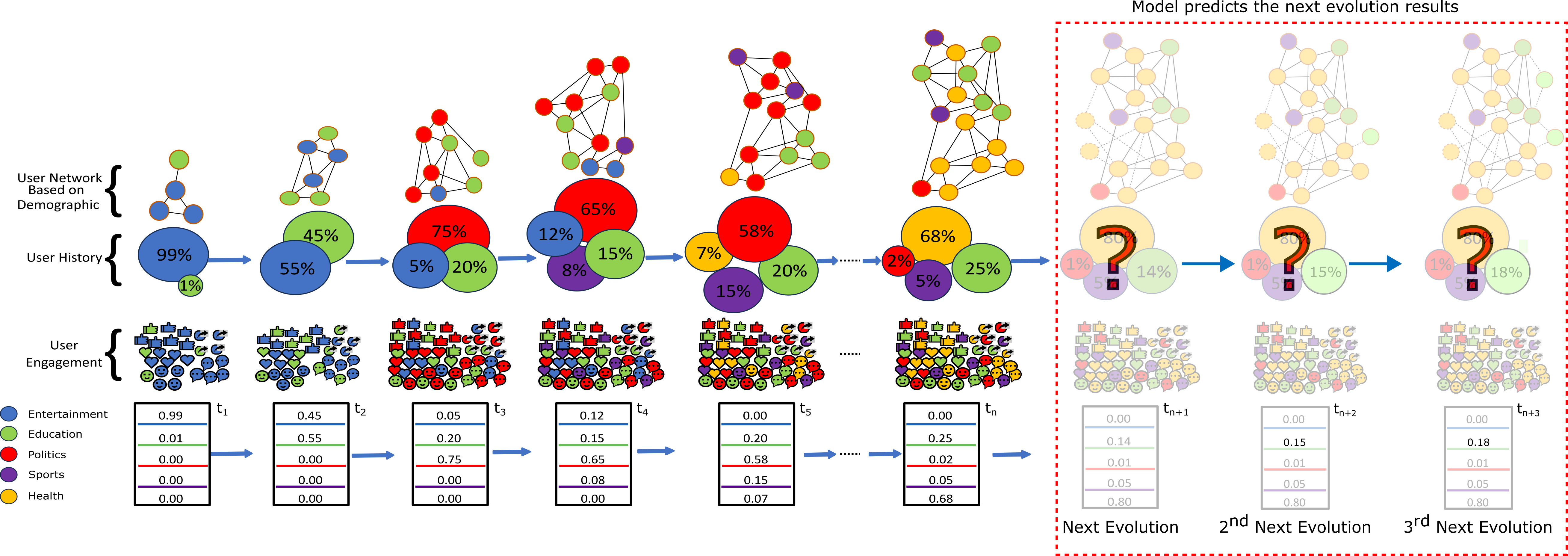}
    \caption{A visualization of user evolution on social media over time}
    \label{fig:system-architecture}
\end{figure*}

Recent advancements in natural language processing have inspired researchers to leverage GPT and BERT architectures, fine-tuning them for tasks beyond text generation. As a decoder-only model, GPT excels in predicting the next item in a sequence~\cite{radford2019language}, enabling its application in tasks such as translation, summarization, and question answering~\cite{radford2019language}. For this study, we utilize a GPT-like decoder-only model~\cite{hossain2024evolve} to predict user social evolution, adapting its proficiency in sequence prediction to forecast changes in user behavior and network dynamics. Inspired by the success of these models, we investigate how user behavior on social media evolves over time. Figure~\ref{fig:system-architecture} illustrates the evolution of a user's social media presence, showing how their posting habits and network connections change. For instance, a user initially focused on entertainment posts (99\%) and rarely engaging with educational content (1\%) might shift over time to increased interest in health and education, with reduced attention to sports and politics. This dynamic evolution mirrors the sequential nature of text generation, allowing us to model user activity analogously to text sequences.

Our goal is to forecast the future state of a user's social media presence, encompassing network dynamics and anticipated activities. Specifically, we predict changes in social connections, such as new friendships or interactions, and future behaviors, including the likelihood of posting specific content. By accurately forecasting these elements, we aim to provide valuable recommendations that enhance user engagement and improve the overall social media experience. To achieve this, we experiment with Llama-3-Instruct, Mistral-7B-Instruct, Gemma-7B-IT using prompt engineering, alongside GPT-2 and BERT with joint feature embedding techniques, to predict the next stages of user evolution. Our experiments demonstrate the effectiveness of this approach, with GPT-2 achieving the lowest perplexity (8.21) in a Cross-modal configuration, outperforming RoBERTa (9.11) and BERT. These results emphasize the importance of leveraging Cross-modal configurations for superior performance, highlighting their potential to predict and analyze user behavior effectively.

In summary, we introduce the following contributions:
\begin{itemize}
    \item \textbf{Modeling Social Evolution Using Demographics, History, and Engagement:} We propose prompt engineering and joint feature embedding techniques that predict online social evolution for a user based on their activities and interactions within social media.

    \item \textbf{The Use Case of Joint Feature Embedding Technique:} With the help of joint embedding technique, we introduce the combining of multiple features of user data to fit into the model to perform the prediction of user evolution on social media.

    \item \textbf{Utilizing LLM for User Evolution Prediction:} We use Llama-3-Instruct, Mistral-7B-Instruct, Gemma-7B-IT to generate predictions about the future state of a user on social media, considering user networks and various attributes.
        
    \item \textbf{The Use Case of Prompt Engineering:} With the help of role-based prompts, we introduce the format of the input for Llama-3-Instruct, Mistral-7B-Instruct, Gemma-7B-IT that drives the models to perform the prediction of user evolution on social media.

    \item \textbf{Proposing a UI for Better User Experience:} We propose a prototype of the user interface to enhance user-level experience as a use case of user evolution on social media.
\end{itemize}

\subsection{Research Objectives} \label{subsec:rq}
In this article, we investigate the following research questions (RQs) related to user evolution in social media. Each question is designed to explore different facets of the problem and contribute to the overarching goal of predicting user behavior effectively.

\begin{itemize}
    \item
    \textbf{(RQ1)}: Can we propose a system that will predict user evolution in social media?
    

    \item
    \textbf{(RQ2)}: Can the system measure the evolution in terms of user data that drives this evolution?
    

    \item
    \textbf{(RQ3)}: How effective is the LLMs, especially Llama-3-Instruct, Mistral-7B-Instruct, Gemma-7B-IT, at accurately predicting user evolution in social media?
    

    \item
    \textbf{(RQ4)}: How effective is the joint feature embedding technique, at accurately predicting user evolution in social media, particularly when the data is ambiguous or multifaceted?


    \item
    \textbf{(RQ5)}: What challenges and opportunities arise when implementing dynamic changes to predict evolution?
    

    \item
    \textbf{(RQ6)}: What are the real-world applications of predicting user evolution in social media?
    
\end{itemize}

Our ultimate goal with this research is to develop a system capable of identifying user evolution in social media, which can be leveraged for important applications such as recommendation systems, enhanced user engagement strategies, and improved content delivery mechanisms. By addressing these research questions comprehensively, we aim to contribute valuable insights and methodologies to the field of social media analytics

\section{Background}
Understanding user evolution in social media requires a multidisciplinary perspective that combines social network theory, behavioral analytics, and machine learning. Social media platforms like Facebook, Twitter, and Reddit are dynamic ecosystems where user behaviors—such as posting frequency, interaction patterns, and network formation—continuously evolve over time. This evolution is influenced by multiple factors including demographic attributes, changes in user interests, peer influence within the network, and external societal events.

Previous research has highlighted the importance of examining temporal shifts in user behavior and network structures. Studies on community evolution have shown that users often transition between communities based on evolving interests~\cite{tracking_dakiche_2019}, while others have focused on how user influence and social status change over time~\cite{analyzing_lin_2009}. The phenomenon of triadic closure, where friends of friends are likely to become connected, also plays a critical role in shaping user evolution on social networks~\cite{role_weng_2013}. Furthermore, the cultural diversity within social platforms and its impact on user dynamics has been explored, emphasizing the importance of demographic and contextual factors~\cite{evolution_grabowski_2006}.

From a modeling standpoint, capturing user evolution requires handling sequential data that reflects interactions, activities, and network changes over time. Natural language processing (NLP) models, particularly sequence prediction models like GPT~\cite{radford2019language}, have demonstrated strong capabilities in forecasting the next elements in a sequence, making them well-suited for predicting user behavioral trajectories in social media. Recent advances, such as the adaptation of decoder-only architectures for non-textual sequential data, further motivate the application of these models in understanding and forecasting user evolution.

In our work, we integrate user demographic profiles, historical activity records, engagement metrics, and network connections to form a comprehensive representation of a user's social media presence over time. By leveraging joint feature embedding techniques and prompt engineering strategies, we aim to predict how users are likely to evolve in their activities and networks. This predictive framework not only provides deeper insights into user behavior but also opens up opportunities for enhancing user experience through personalized recommendations, targeted community-building efforts, and adaptive platform design.

In evaluating the prediction of user evolution, it is crucial to select metrics that align with the sequential and categorical nature of the task. We adopt Perplexity and Pseudo-Perplexity to measure how well the model can predict the next state of user behavior, as lower perplexity values indicate stronger sequence modeling capability. Additionally, we utilize Precision@10 (PR@10) and Hits@10 to assess the quality of predicted future user activities and connections, focusing on the top-10 most likely outcomes.

\section{Literature Review} \label{sec:lr}
User evolution in social media refers to how users' behaviors, preferences, and interactions change over time. Prior studies have examined various aspects of this evolution, including engagement patterns, content creation, and network dynamics.


\textbf{Engagement Growth Phase.}
As users get accustomed to Twitter, they follow more accounts, actively engage with tweets, and join trending topics using hashtags, becoming active participants~\cite{Joinson2008}. They create more content, share opinions, and enhance their profiles with detailed bios and pictures~\cite{Lampe2006}. Their follower count grows with increased interaction, boosting visibility~\cite{Burke2009}.

\textbf{Personalization and Customization Phase.}
Users personalize their feed by following relevant accounts and muting or unfollowing those that are not~\cite{Kwak2010}. Engagement grows selective, with users joining interesting discussions and sharing significant content like threads or media-rich posts~\cite{Li2014}. They refine profiles to showcase their brand or identity, pinning key tweets and using cover photos~\cite{McCay-Peet2017}.

\textbf{Influence and Authority Phase.}
Users evolve into influencers, gaining a large following and high engagement, which is crucial for shaping online communities and trends~\cite{Marwick2011}. Their content becomes strategic to maintain and grow their audience. Influencers collaborate with brands, users, and engage in high-profile discussions~\cite{Cha2010}, influencing trends and conversations on the platform~\cite{Jin2014}.

\textbf{Mature User Phase.}
Mature users post less frequently but stay consistently active, prioritizing quality and meaningful interactions~\cite{Bakshy2011}. They follow trusted sources and engage with content that aligns with their interests and values~\cite{Brandtzaeg2009}. They contribute to community building by supporting new users, joining discussions, and moderating communities~\cite{Tang2012}.

\textbf{Implications for Recommendation Systems.}
As users evolve, recommendation systems must adapt to their changing preferences. Initial recommendations focus on onboarding content, while later ones become more personalized~\cite{Kumar2010}. Understanding user evolution is key to predicting future engagement and suggesting content aligned with current interests and trends~\cite{Ricci2015}. Analyzing network changes over time aids in recommending new connections or communities of interest~\cite{jannach2010recommender}.

\section{Problem Formulation} \label{sec:problem-formulation}
To address \textbf{RQ2}, we designed our system based on user data that drives user evolution in social media. Below, we design the detail formulation process of our proposed system.
Let $U$ be the set of users and $A$ be the adjacency matrix that represents the connections between users if they are friends or followers on social media. Demographic data can be defined by $D$, where $D = \{a, g, o, l\}$, with $a$ representing age, $g$ representing gender, $o$ representing occupation, and $l$ representing location. Let $H$ represent user history based on post categories $C = \{c_1, c_2, c_3, \ldots, c_k\}$, and $E$ represent user engagement, which includes interactions with posts such as reactions, shares, and comments. The feature vector $X$ consists of demographic information, user history, and user engagement. Suppose we have $T$ time steps for $X$ for a single user.

We then create a combined feature matrix for each time step $t \in \{1, 2, \dots, T\}$. One approach is to concatenate the adjacency matrix, demographic matrix, history matrix, and engagement matrix for $N$ users to produce the feature embedding $F$.

\[
F^{(t)} = \left[ A^{(t)} \| D^{(t)} \| H^{(t)} \| E^{(t)} \right]
\]

\[
A^{(t)}_{ij} = 
\begin{cases}
    1 & \text{if a connection exists between users $u_i$ and $u_j$} \\
    0 & \text{otherwise}
\end{cases}
\]

\textbf{Input:} Construct a sequence of combined feature matrices over the $T$ time steps: $S = \left[ F^{(1)}, F^{(2)}, F^{(3)}, \dots, F^{(t)}, F^{(t+1)}, \dots, F^{(T)}\right]$

\textbf{Output.}\;
An evolution model that estimates the conditional probability
\begin{align}
  &P\!\Bigl(
        F^{(T+1)},\,F^{(T+2)},\,F^{(T+3)},\,F^{(T+4)}
     \,\Big|\,
        F^{(1)},\,F^{(2)},\,\dots,\,F^{(T)}
    \Bigr) \\
  &\text{evaluated at the steps } (T+1),\,(T+2),\,(T+3),\text{ and }(T+4).
\end{align}

\section{Motivation} \label{sec:motivation}
Social media platforms are crucial for shaping identities. Understanding user behavior is key for improving experiences and informing business strategies. Analyzing user trajectories reveals insights into engagement, preferences, and social connections.
Forecasting a user's social media future helps platforms offer personalized experiences matching evolving interests. As users change through life stages like education or careers, their interests and networks shift. Accurately predicting these changes improves user satisfaction and engagement with relevant content and connections.
Understanding user trajectories helps identify trends and behaviors that inform new feature development. For instance, if many users engage with certain content, platforms can use this to build communities and promote similar content, enhancing user retention and growth.

We investigate predicting social media network dynamics and activities by analyzing user history to forecast future actions and connections \textbf{RQ1}. Social media evolution parallels text sequences, influenced by interests, education, and professional networks. We use Llama-3-Instruct, Mistral-7B-Instruct, Gemma-7B-IT, GPT-2, and BERT to foresee changes in user networks, providing insights for users and developers. Initially, users engage with educational content, then shift to industry-related interests and networking as they start their careers. Using these models, including RoBERTa, we predict future behavior by examining past activities.

\section{Methodology} \label{sec:methodology}

\subsection*{Temporal Notation} 
Let $t \in \{1,\dots,T\}$ denote discrete time-steps (e.g.\ weeks) and $t\!+\!\Delta$ the prediction horizon.
All graph snapshots $A^{(t)}$, features $D^{(t)}$, histories $x_P^{(t)}$, etc.\ are indexed accordingly. 

\begin{figure}[h]
    \centering
    \includegraphics[width=0.45\textwidth]{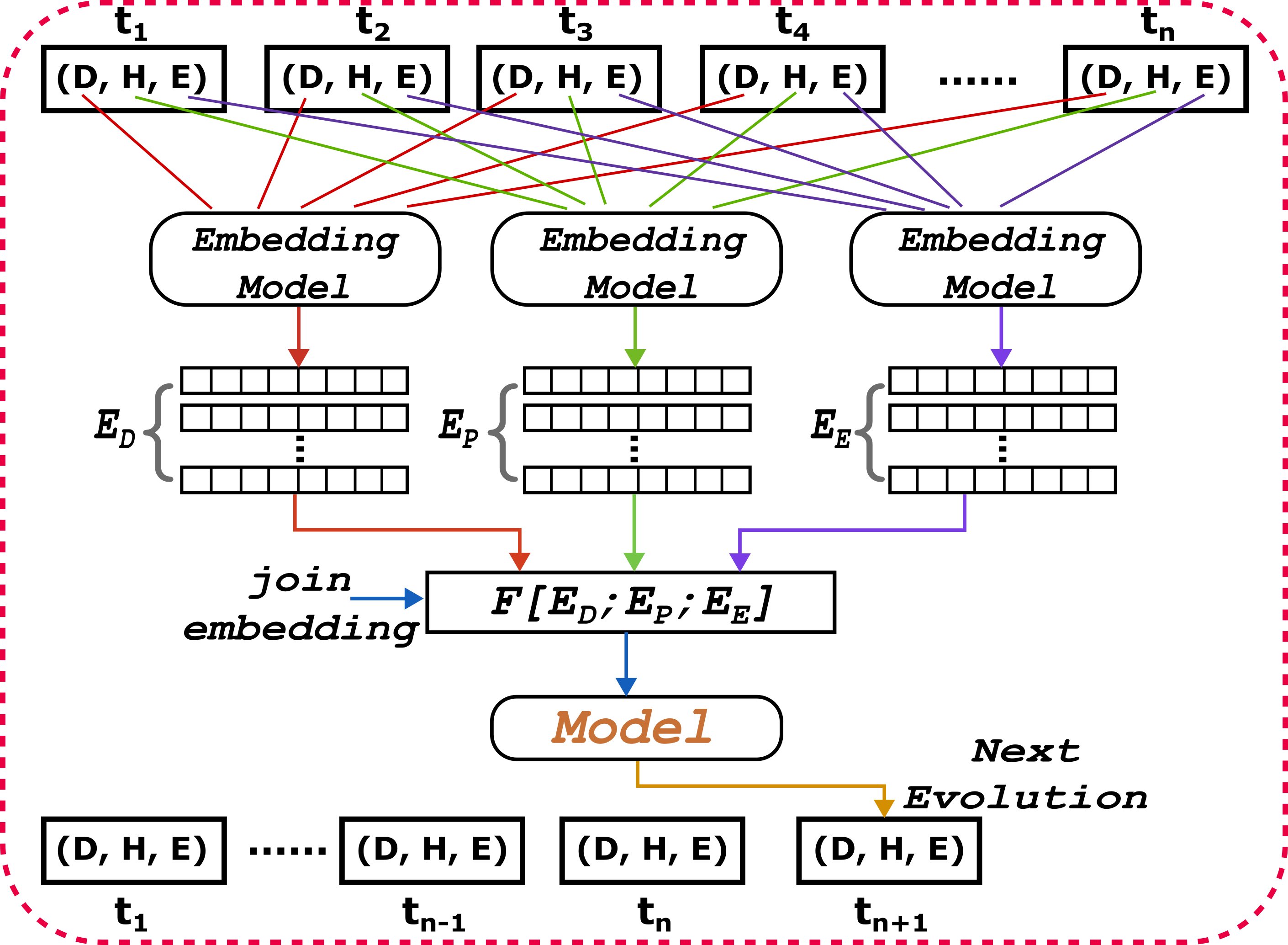}
    \caption{User Next Evolution Prediction using Join Embedding method.}
    \label{fig:join-embedding}
\end{figure}

\subsection{Input Representations}
\begin{itemize}
    \item \textbf{Demographic Features:} By combining the adjency matrix, $A^{(t)}$ and demographic features, $D^{(t)}$, we create $x_D$.
    Let $x_D \in \mathbb{R}^{n_D}$ represent demographic features.
    \[
    E_D = f_D(x_D) \in \mathbb{R}^{d_D}
    \]
    Here, $E_D$ is the demographic feature embedding and $n_D$ is the dimension of the feature embedding.
    Here, $f_D$ represents Graph Neural Network (GNN) to create the demographic feature embedding, $E_D$ to capture the structural relationships between nodes (users) in the graph (user-network). 
    
    \item \textbf{Post History Features:} Let $x_P = \{p_1, p_2, \dots, p_n\}$ represent the sequence of $n$ posts at time $t$.
    \[
    E_P = f_P(x_P) \in \mathbb{R}^{d_P}
    \]
    In this context, $f_P$ represents a pre-trained model such as BERT, RoBERTa, GPT-2. We generate the post embedding $E_P$ by leveraging the BERT, RoBERTa, and GPT-2 model. To develop the feature embedding, we exclusively use the textual content of a post, which is subjected to data pre-processing comprising data cleaning and tokenization.
    
    \item \textbf{Engagement Features:} Let $x_E \in \mathbb{R}^{n_E}$ denote the metrics of engagement. The vector $x_E$ includes user data on reactions, comments, and shares categorized by type, as previously outlined in this study. For instance, at time $t$, a user might have the following counts for the politics category: {'reactions': 100, 'comments': 50, 'shares': 20} out of 1000 posts. These categories align with posts where the user has shown any form of engagement, be it a reaction, comment, or share. We have created user engagement summary (as described in Appendix A~\ref{subsec:appendixa}) in order to create embedding by utilizing BERT, RoBERTa, and GPT-2.
    \[
    E_E = f_E(x_E) \in \mathbb{R}^{d_E}
    \]
    Here, $E_E$ is the engagement feature embedding and $d_E$ is the dimension of this embedding.
\end{itemize}

\subsection{Fusion of Embeddings}
We have demographic, post history and engagement history feature embeddings $E_D, E_P, E_E$. Now, we are going to join or combine these three embedding to have the new embedding. This combined embedding will be using to train the model for final prediction of the next user evaluation. We perform embedding joining by apply three different approaches. Figure~\ref{fig:join-embedding} demonstrates the join embedding method with given user history and prediction the next evolution. 
\begin{itemize}
    \item \textbf{Concatenation:} Concatenation combines multi-modal embeddings by stacking them, effectively integrating diverse data without extra parameters. It's effective when modalities provide independent, complementary information. This method takes all embeddings and concat these to have a single embedding, $F$.
    \[
    F = [E_D; E_P; E_E] \in \mathbb{R}^{d_D + d_P + d_E}
    \]
     In social media user modeling, demographics like age and profession help with user segmentation, while engagement data gives immediate behavioral insights. Studies such as~\cite{ngiam2011multimodal} show that concatenation is a strong baseline in multi-modal tasks, though it may struggle with capturing complex interactions between modalities.

    \item \textbf{Attention-Based Fusion:}
    Attention-based fusion dynamically prioritizes modalities based on task relevance, unlike static concatenation. When predicting user engagement, it may focus on $E_E$ over $E_D$ for active users, and $E_D$ for long-term trends. Instead of directly adding all embedding it multiply weights $\alpha_D, \alpha_P, and \ \alpha_E$ to the $E_D, E_P, and E_E$ respectively and add them to have final embedding, $F$.
    \[
    \alpha_D = \text{softmax}(W_D E_D), \quad \alpha_P = \text{softmax}(W_P E_P),
    \]
    \[
    \alpha_E = \text{softmax}(W_E E_E)
    \]
    where $W_D \in \mathbb{R}^{1 \times d_D}$, $W_P \in \mathbb{R}^{1 \times d_P}$, and $W_E \in \mathbb{R}^{1 \times d_E}$.

    \[
    F = \alpha_D E_D + \alpha_P E_P + \alpha_E E_E
    \]
     Introduced by \cite{vaswani2017attention}, attention mechanisms are vital in multi-modal learning, offering improved performance and interpretability through learnable weights and scores. As \cite{tsai2019multimodal} demonstrates, they enhance multi-modal sentiment analysis by highlighting key modalities.
    \item \textbf{Cross-Modal Attention:}
    Cross-modal fusion goes beyond individual modalities by focusing on their interactions to reveal complex dependencies. Embeddings interact through mechanisms like cross-attention or bilinear transformations. For instance, it models how demographics ($E_D$) affect user engagement ($E_E$) with content categories ($E_P$), offering deeper user behavior insights.
    \[
    Q = W_Q F_{\text{prev}} \in \mathbb{R}^{d_q}, \quad K = [E_D, E_P, E_E] \in \mathbb{R}^{d_k \times 3},
    \]
    \[
    V = [E_D, E_P, E_E] \in \mathbb{R}^{d_v \times 3}
    \]
    \[
    F = \text{softmax}\left(\frac{Q K^T}{\sqrt{d_k}}\right) V
    \]
     This approach excels in tasks needing contextual relationship understanding, as shown in \cite{sun2019videobert}, where it was crucial in visual question answering. By focusing on specific modality pairs or triads, it captures nuanced relationships overlooked by static methods like concatenation. In social media, this might involve dynamically assessing how education level influences engagement with tech-related posts in real time.
\end{itemize}

\subsection{Prediction}
The fused embedding $F$ is passed through a feed-forward layer:
\[
y = f_{\text{predictor}}(F)
\]
where $y$ is predicted results. $y$ is then utilized for further predictions: link prediction among the users and the category-wise user activity prediction.

\subsubsection{Link Prediction}
Link prediction aims to predict whether an edge exists between two nodes (users). In this task, the model predicts the probability of an edge between a pair of nodes \( (i, j) \).

\textbf{Step 1}: To predict a link, we first extract the combined embeddings for node pairs \( i \) and \( j \), denoted as \( \mathbf{y}_i \) and \( \mathbf{y}_j \), respectively.

\textbf{Step 2}: These embeddings are combined to form a joint representation of the node pair:
\[
\mathbf{y}_{ij} = \text{concat}(\mathbf{y}_i, \mathbf{y}_j)
\]

\textbf{Step 3}: The resulting joint embedding \( \mathbf{y}_{ij} \) is passed through a \textbf{classifier} (e.g., a fully connected layer) to predict the probability of an edge between the nodes:
\[
P(\text{edge}(i, j)) = \sigma(\mathbf{y}_{ij})
\]
where \( \sigma \) is the sigmoid activation function.

\textbf{Step 4}: A threshold, $\theta = 0.5$ is applied to the output of the classifier to determine whether an edge exists:
\begin{itemize}
    \item If \( P(\text{edge}(i, j)) > \theta \), predict an edge exists between nodes \( i \) and \( j \).
    \item If \( P(\text{edge}(i, j)) \leq \theta \), predict no edge exists between nodes \( i \) and \( j \).
\end{itemize}

\subsubsection{User Activity Prediction}
User Activity Prediction involves predicting the liklihood of each category, $c_k$ for each node (user) based on its embeddings. Given the predicted output \( \mathbf{y}_i \) of $f_{predictor(.)}$ for each node \( i \), we use a \textbf{sigmoid} function, $\sigma$ to predict the probability of each category.

The output is calculated as:
\[
P([c_1, c_2, c_3, \cdots, c_k]) = \sigma(\mathbf{y}_i)
\]
where \( P[...]\) is the probability value of all categories.

\subsubsection{Training Procedure}
BERT, RoBERTa and GPT-2 are fine-tuned. The loss function for training consists of two components:
\begin{itemize}
    \item \textbf{Link Prediction Loss}: Binary cross-entropy loss for link prediction tasks:
    \begin{align*}
    \mathcal{L}_{\text{link}} &= - \sum_{(i,j) \in \mathcal{E}} A^{(t)}_{ij} \log(P(\text{edge}(i,j))) \\
    & \quad + (1 - A^{(t)}_{ij}) \log(1 - P(\text{edge}(i,j)))
    \end{align*}
    where \( A^{(t)}_{ij} \) is the true label for the edge (1 if an edge exists, 0 otherwise) at time $t$, and \( P(edge(i,j)) \) is the predicted probability for the edge.
    
    \item \textbf{User Acitivity Prediction Loss}: Categorical cross-entropy loss for user activity prediciton tasks:
    \[
    \mathcal{L}_{\text{activity}} = - \sum_{i \in \mathcal{V}} \sum_{c \in \mathcal{C}} \frac{x_{ic}}{\max(x_i)} \log(P_{ic})
    \]
    where \( x_{ic} \) is the true value for category \(c\) for node (user) \( i \), \( x_{i} \) is true count of all categories for user \(i\), and \( P_{ic} \) is the predicted probability for category \( c \).
    
    Needless to say, as in our dataset, we have the count of each category at time, $t$ for a user like-- \{`Polictics': 200, `Education': 50, `Sports': 20, `Travel': 5, ....\} in terms of user post history. So, we normalize them to be in the range \([0, 1]\) to align them with the predicted probabilities and that is being done by \( \frac{x_{ic}}{\max(x_i)}\).
    
\end{itemize}

The total loss function is the weighted sum of these two components:
\[
\mathcal{L} = \lambda_2\mathcal{L}_{\text{link}} + \lambda_2 \mathcal{L}_{\text{activity}}
\]
where \( \lambda_1, \lambda_2 \) are two regularization parameter that balances the two losses. We evaluate our model with different value of \( \lambda_1, \lambda_2 \) (as shown in Table~\ref{tab:model-performance-table-1} and Table~\ref{tab:model-performance-table-2}).

The description about Model Evaluation Metrics for link and user activities prediction is given in the Appendix A~\ref{subsec:appendixa}.

\subsection{Prompt Engineering in Predictive User Evolution Using LLM}
Prompt engineering is a crucial aspect of leveraging language models for specific tasks. By designing tailored prompts, we can guide the model to generate more accurate and relevant outputs. This section delves into the concept of role-based prompts and illustrates their application in predicting user evolution based on historical data.

\textbf{Role-Based Prompts.}
Role-based prompts are a structured form of prompt engineering where the model is assigned a specific role, such as a data scientist, software engineer, or market analyst. This role provides context and expectations, which in turn helps the model generate responses that are more aligned with the requirements of the task. By clearly defining the role, task, context, data, and instructions, role-based prompts ensure that the model's responses are focused and relevant.

\textbf{Use Case: Predicting User Evolution.}
In our use case, we aim to predict the next evolution of user activities which include user engagement, interests, and demographics based on the user histoy. The data is segmented into four main categories: User Graph, User History, User Engagement Scores, and User Demography. To achieve this, we employ a role-based prompt designed for a data scientist. The prompt provides a comprehensive framework that includes detailed instructions, contextual information, and example outputs. To address the \textbf{RQ3}, We evaluated with the role base prompt structure, to show how effective the open-source llm like Llama-3, Mistral-7B, and Gemma-7B.

\begin{figure}[h]
    \centering
    \includegraphics[width=0.3\textwidth]{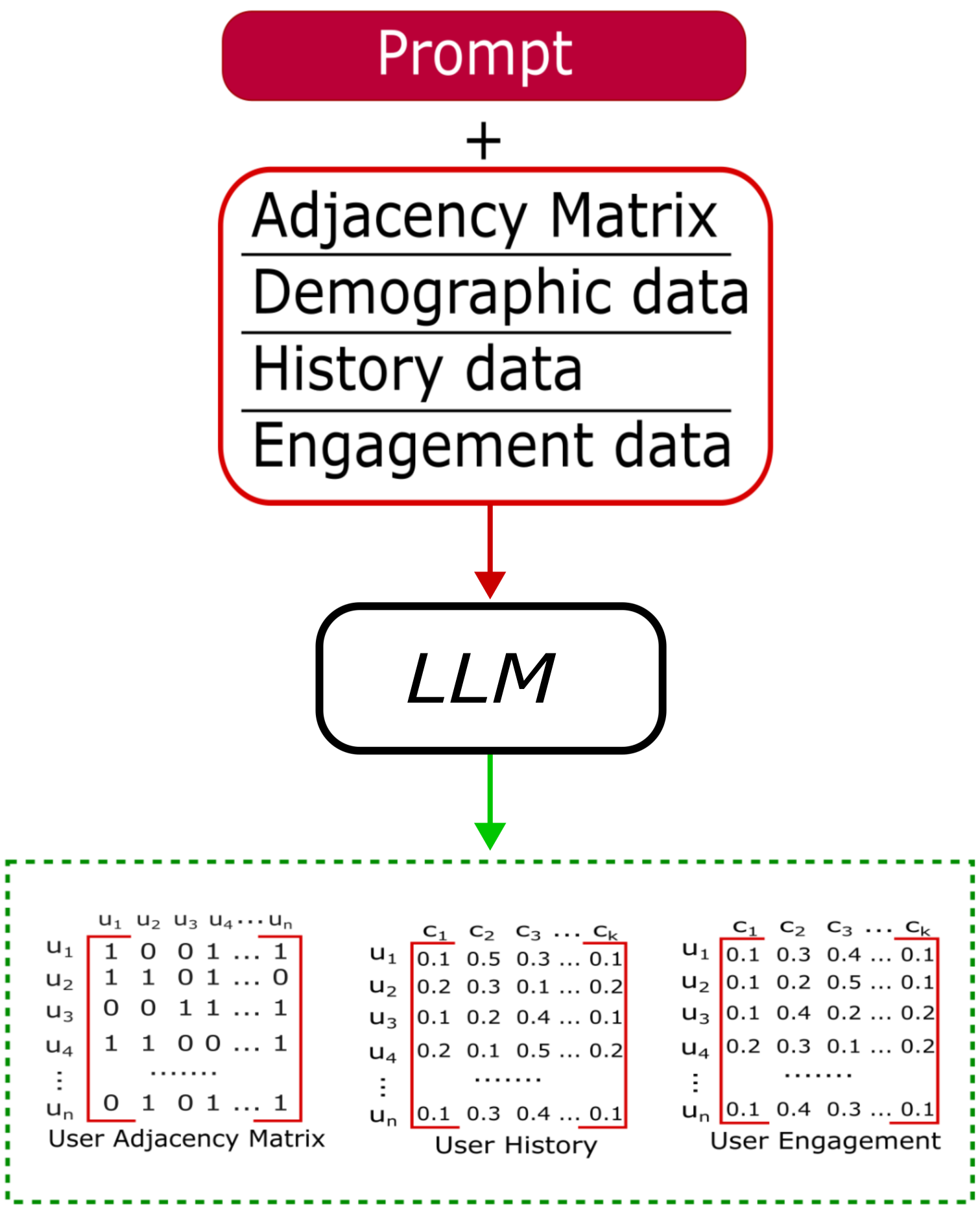}
    \caption{User Next Evolution prediction using LLM}
    \label{fig:gpt4o-evolution}
\end{figure}

\section{Experiment} \label{sec:experiment}

\subsection{Dataset} \label{subsec:dataset}
To study user evolution on social media, we compiled a dataset of 5,000 Facebook posts from 500 purposively sampled users, including friends, mutual connections, public figures, celebrities, athletes, and journalists. 
% ...
After completing data collection and preprocessing, we successfully compiled a dataset of 5,000 Facebook posts along with 500 users who have been active for five years or more.
We have \textbf{72\%} users with five–seven years of Facebook activity 
and \textbf{65\%} users whose activity spans one–five categories. 
% ...

\textbf{Ethical compliance.}  
All data were obtained exclusively from public Facebook pages in strict accordance with the platform’s rules and Terms of Service; no friend-only or private content was retrieved. User identifiers were irreversibly hashed and all records were stored on encrypted drives, ensuring that neither the researchers nor downstream models could recover personal identities. The complete collection and anonymisation protocol was reviewed and approved by the university’s Institutional Review Boards.


\section{Experimental Setup}

\textbf{Data Preprocessing.}
The text data are preprocessed 
by tokenization, lower-casing, and optional stop-word removal.
Demographic \emph{numeric} features (e.g., age) are z-score normalized.
Categorical attributes (profession, location) are \emph{not} z-scored; they are encoded via one-hot vectors or learned embeddings. 
Engagement counts are min-max scaled.
% ...

\textbf{Tokenization.}
For BERT and RoBERTa we use the \texttt{BertTokenizer} from Hugging Face. 

\textbf{Hyperparameters.}
We follow BERT/RoBERTa and GPT-2. 
Fine-tuning 
uses a learning rate of $2\!\times\!10^{-5}$ for BERT/RoBERTa and $1\!\times\!10^{-5}$ for GPT-2, with linear decay.
% ...

\begin{table}
\centering
\caption{Model performance comparison in terms of evaluation metrics, Pseudo-Perplexity (PPLL), Perplexity (PPL), when $\lambda_1=0.3, \lambda_2=0.7$, (* denotes that model evaluated by PPL).} 
\label{tab:model-performance-table-1}
    \begin{tabular}{c|cccc}
        \hline
         Model & Concat & Attention & Cross-modal \\
         \hline
         RoBERTa & 18.35 & 15.77 & 9.79 \\
         BERT & 21.13 & 19.62 & 16.63 \\
         $GPT-2^*$ & 16.68 & 14.19 & 7.61 \\
         \hline
    \end{tabular}
\end{table}

\begin{table}
\centering
\caption{Model performance comparison in terms of evaluation metric, Pseudo-Perplexity (PPLL), Perplexity (PPL), when $\lambda_1=0.5, \lambda_2=0.5$, (* denotes that model evaluated by PPL).}
\label{tab:model-performance-table-2}
    \begin{tabular}{c|cccc}
        \hline
         Model & Concat & Attention & Cross-modal \\
         \hline
         RoBERTa & 16.95 & 16.47 & 9.11 \\
         BERT & 20.19 & 17.37 & 14.59 \\
         $GPT-2^*$ & 14.51 & 13.33 & 8.21 \\
         \hline
    \end{tabular}
\end{table}

\begin{table}[t]
\centering
\caption{Instruction-tuned LLM performance when $\lambda_1=0.4,\;\lambda_2=0.6$ (higher is better).}
\label{tab:model-performance-table-4}
\begin{tabular}{lccc}
\hline
Model & AUC-ROC (Link) & Hits@10 (Link) & Macro-F1 (Activity)\\
\hline
Llama-3-Instruct$^{*}$     & 0.91 & 0.59 & 0.78\\
Mistral-7B-Instruct$^{*}$  & 0.88 & 0.55 & 0.74\\
Gemma-7B-IT$^{*}$          & 0.85 & 0.52 & 0.71\\
\hline
\end{tabular}
\end{table}


\subsection{Results and Analysis}
\noindent
Tables~\ref{tab:model-performance-table-1} and~\ref{tab:model-performance-table-2} address \textbf{RQ4} by comparing RoBERTa, BERT, and GPT-2 under varying regularisation balances ($\lambda_{1}$, $\lambda_{2}$) and three fusion strategies—concat, attention, and cross-modal attention.  In every configuration the cross-modal variant delivers the lowest pseudo-perplexity or perplexity: RoBERTa drops to 9.79/9.11\,PPLL compared with 15.77/16.47 for vanilla attention and 18.35/16.95 for concatenation; BERT follows the same pattern (16.63/14.59 versus 19.62/17.37 and 21.13/20.19).  GPT-2 remains the overall leader, achieving 7.61/7.21\,PPL with cross-modal attention.  These results show that explicitly aligning textual and behavioural signals benefits both masked (RoBERTa, BERT) and autoregressive (GPT-2) models.

\smallskip
\noindent
Table~\ref{tab:model-performance-table-4} extends this finding to larger instruction-tuned LLMs trained with the multi-objective regime of Section~3 ($\lambda_{1}=0.4$, $\lambda_{2}=0.6$).  Using four heterogeneous user signals—demographics, post history, engagement statistics, and network edges—each model is evaluated on link prediction (\textit{AUC-ROC} and \textit{Hits@10}) and user-state classification (\textit{Macro-F1}).  Llama-3-Instruct leads with 0.91, 0.59, and 0.78, respectively, while Mistral-7B-Instruct trails by two to three points and Gemma-7B-IT shows the steepest drop.  The consistent ordering across metrics supports our claim that richer linguistic priors and a balanced loss (\,$\lambda_{1}$, $\lambda_{2}$\,) enhance graph reasoning.```

\section{Application}
\label{sec:application}

\begin{figure*}[h]
    \centering
    \includegraphics[width=\textwidth]{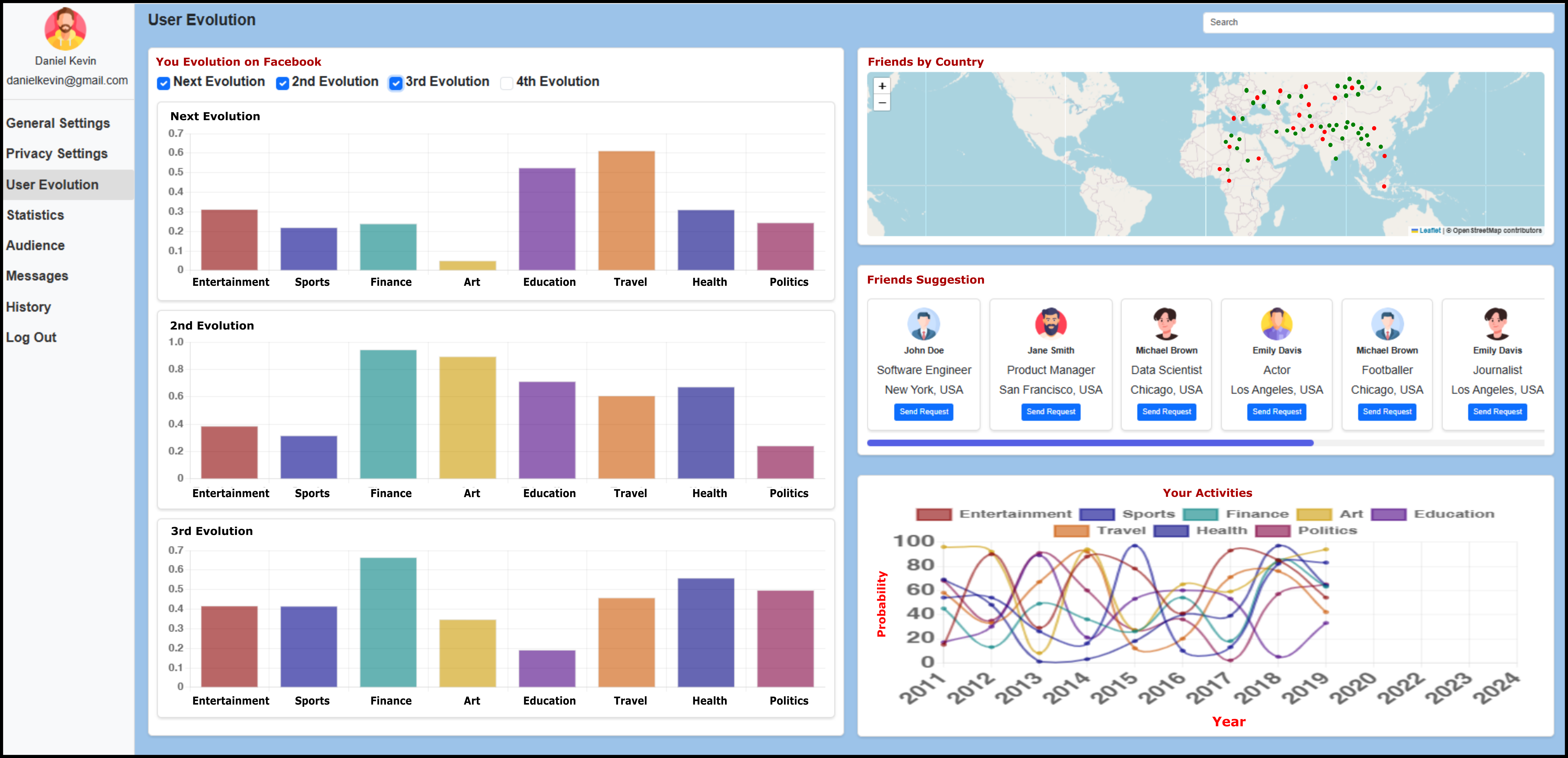}
    \caption{User Interface depicting the next and subsequent evolution stages of a user on social media}
    \label{fig:ui}
\end{figure*}

In addressing \textbf{RQ6}, a primary application of this study is recommending future friends to users as part of their network growth, i.e., predicting future connections within the same network. Additionally, this study anticipates user activities related to social media posting and their likelihood of engaging with other users' posts. User behaviors and potential future connections are depicted across four stages of evolution. Figure~\ref{fig:ui} showcases three consecutive stages: Next Evolution and Second Evolution and Third Evolution. Users can also explore fourth evolution stages. Needless to say, this system gives an overview of the users' future direction in their activities on social media. This podcast of future activities can warn users to stop activities that can bring harm to self even to others. For example, If a user is becoming more extremist and friends are suggested for the next evolution might have a negative impact. So, before becoming friends or followers of others users have a chance to check about them. We have three other visulations in the UI-- Friends by Country, Friends Suggestion, and User Activities.

\textbf{Friends by Country.} It shows a map and some green/red dots on the map. Green dots indicate that users who connect with the user as either friends or followers and Red dots for who are the potential users in the same network would be the friends or follower in future. Positions of the red points would change based the which evolution stage is checked on the UI.

\textbf{Friends Suggestion.} In the part those who are potential users as like the friends suggestion
happens on the social media, but it is different than in terms of user selection of evolution stage
on the UI. The suggestion is rearranged by the user activites on different categories.

\textbf{User Activities.} It describes the user activities from the begining of the present time on 
different categories. With eight different lines, the graph depicts that over the time how user 
activities change.

\section{Discussion and Limitations} \label{sec:discussion}
Our user evolution forecast model is capable of predicting user activities and user networks for the next four stages. However, to address, \textbf{RQ5}, using the predicted result of one stage to predict others has been shown to reduce performance. This drop in performance may result from assigning random values to missing data and the size of the dataset. Additionally, we have not included other demographic attributes such as race, ethnicity, religion, hobbies, and marital status in the dataset, which could potentially affect the overall model performance. On top of that, we specify the categories to total eight which might not be proper set of categories, there might have more categories instead. Nonetheless, the system built with our prediction model is able to recommend friends based on the evolution stage, predicting potential future friends or followers. This capability helps users avoid potentially vulnerable connections based on factors such as location and profession. Users will be consistently informed about activities on social media, aiding them in limiting interactions with such vulnerable friends and followers.

\section{Ethical Considerations}
\label{sec:ethical}

To ensure user privacy and uphold ethical standards, strict anonymization protocols were implemented throughout the research. Personally identifiable information (PII), such as usernames, locations, and other sensitive attributes, was systematically removed from the dataset. The anonymized data was stored using unique user identifiers that were not linked to any PII, ensuring confidentiality and untraceability. These practices align with established privacy standards, including NIST SP 800-122~\cite{NIST} and the General Data Protection Regulation (GDPR)~\cite{GDPR}, guaranteeing that all data processing complies with legal and ethical requirements.

This study used anonymized datasets while prioritizing user agency and informed consent. Data collection and processing followed ethical guidelines, safeguarding user rights and privacy. The framework emphasized transparency, ethical handling, user empowerment, and responsible usage. Though anonymization limited direct user involvement, ethical research was conducted respecting user dignity and data sovereignty. This study's measures prioritize privacy, fairness, and transparency by preventing re-identification, following ethical standards, and ensuring responsible data handling to reduce risks in predictive modeling. These efforts ensure respect for individual privacy and address societal concerns about predictive technologies.

\section{Conclusion}
\label{sec:conclusion}
In this study, we have elucidated the process of online social evolution and pinpointed the primary attributes driving this phenomenon. we experimented with the open-source llm like models Llama-3-Instruct, Mistral-7B-Instruct, Gemma-7B-IT, leveraging its capabilities akin to next-word or sentence prediction. Through the analysis of user data across various time steps—including user adjacency matrices, demographic information, history, and engagement—we demonstrated how these evolving datasets propel social evolution. Along with that we leverage GPT-2, BERT, RoBERTa models by utilizing joint feature embeddings to predict user network dynamics and activities on social media.

We presented our findings based on three distinct metrics: Pseudo-Perplexity, Perplexity, Precisions@10, Hits@10, and Accuracy. These results underscore the substantial influence of user attributes in establishing new connections or fortifying existing ones within the network. They also illustrate how changes in user activities drive social media evolution over time. We also developed an UI prototype to show the use case of our model.

\section*{Acknowledgements}
ChatGPT was utilized to assist with language editing and clarity improvements in this work. No content was generated related to technical results, data, code, or analysis.




\begin{thebibliography}{30}


\ifx \showCODEN    \undefined \def \showCODEN     #1{\unskip}     \fi
\ifx \showISBNx    \undefined \def \showISBNx     #1{\unskip}     \fi
\ifx \showISBNxiii \undefined \def \showISBNxiii  #1{\unskip}     \fi
\ifx \showISSN     \undefined \def \showISSN      #1{\unskip}     \fi
\ifx \showLCCN     \undefined \def \showLCCN      #1{\unskip}     \fi
\ifx \shownote     \undefined \def \shownote      #1{#1}          \fi
\ifx \showarticletitle \undefined \def \showarticletitle #1{#1}   \fi
\ifx \showURL      \undefined \def \showURL       {\relax}        \fi
\providecommand\bibfield[2]{#2}
\providecommand\bibinfo[2]{#2}
\providecommand\natexlab[1]{#1}
\providecommand\showeprint[2][]{arXiv:#2}

\bibitem[GDP(2021)]%
        {GDPR}
 \bibinfo{year}{2021}\natexlab{}.
\newblock \bibinfo{title}{General data protection regulation (gdpr)}.
\newblock
\urldef\tempurl%
\url{https://gdpr-info.eu/}
\showURL{%
\tempurl}
\newblock
\shownote{Accessed: 2021-02-12}.


\bibitem[NIS(2021)]%
        {NIST}
 \bibinfo{year}{2021}\natexlab{}.
\newblock \bibinfo{title}{Guide to protecting the confidentiality of personally identifiable information (pii)}.
\newblock
\urldef\tempurl%
\url{https://tinyurl.com/ylyjst5y}
\showURL{%
\tempurl}
\newblock
\shownote{Accessed: 2021-02-12}.


\bibitem[Allamanis et~al\mbox{.}(2012)]%
        {evolution_allamanis_2012}
\bibfield{author}{\bibinfo{person}{Miltiadis Allamanis}, \bibinfo{person}{S. Scellato}, {and} \bibinfo{person}{C. Mascolo}.} \bibinfo{year}{2012}\natexlab{}.
\newblock \showarticletitle{Evolution of a location-based online social network: analysis and models}. In \bibinfo{booktitle}{\emph{Internet Measurement Conference}}.
\newblock


\bibitem[Brandtzæg and Heim(2009)]%
        {Brandtzaeg2009}
\bibfield{author}{\bibinfo{person}{P.~B. Brandtzæg} {and} \bibinfo{person}{J. Heim}.} \bibinfo{year}{2009}\natexlab{}.
\newblock \showarticletitle{Why people use social networking sites}. In \bibinfo{booktitle}{\emph{Online Communities and Social Computing}}. \bibinfo{pages}{143--152}.
\newblock


\bibitem[Burke et~al\mbox{.}(2009)]%
        {Burke2009}
\bibfield{author}{\bibinfo{person}{M. Burke}, \bibinfo{person}{C. Marlow}, {and} \bibinfo{person}{T. Lento}.} \bibinfo{year}{2009}\natexlab{}.
\newblock \showarticletitle{Feed me: Motivating newcomer contribution in social network sites}. In \bibinfo{booktitle}{\emph{Proceedings of the SIGCHI Conference on Human Factors in Computing Systems}}.
\newblock


\bibitem[Dakiche et~al\mbox{.}(2019)]%
        {tracking_dakiche_2019}
\bibfield{author}{\bibinfo{person}{Narimene Dakiche}, \bibinfo{person}{Fatima Benbouzid-Si Tayeb}, \bibinfo{person}{Yahya Slimani}, {and} \bibinfo{person}{Karima Benatchba}.} \bibinfo{year}{2019}\natexlab{}.
\newblock \showarticletitle{Tracking community evolution in social networks: A survey}.
\newblock \bibinfo{journal}{\emph{Information Processing and Management}} (\bibinfo{year}{2019}).
\newblock


\bibitem[et~al.(2011)]%
        {Bakshy2011}
\bibfield{author}{\bibinfo{person}{E.~Bakshy et al.}} \bibinfo{year}{2011}\natexlab{}.
\newblock \showarticletitle{Everyone's an influencer: quantifying influence on twitter}. In \bibinfo{booktitle}{\emph{Proceedings of the Fourth ACM International Conference on Web Search and Data Mining}}.
\newblock


\bibitem[et~al.(2015)]%
        {Ricci2015}
\bibfield{author}{\bibinfo{person}{F.~Ricci et al.}} \bibinfo{year}{2015}\natexlab{}.
\newblock \bibinfo{booktitle}{\emph{Recommender Systems Handbook}}.
\newblock \bibinfo{publisher}{Springer}.
\newblock


\bibitem[et~al.(2010a)]%
        {Kwak2010}
\bibfield{author}{\bibinfo{person}{H.~Kwak et al.}} \bibinfo{year}{2010}\natexlab{a}.
\newblock \showarticletitle{What is twitter, a social network or a news media?}. In \bibinfo{booktitle}{\emph{Proceedings of the 19th International Conference on World Wide Web}}.
\newblock


\bibitem[et~al.(2012)]%
        {Tang2012}
\bibfield{author}{\bibinfo{person}{J.~Tang et al.}} \bibinfo{year}{2012}\natexlab{}.
\newblock \showarticletitle{mtrust: discerning multi-faceted trust in a connected world}. In \bibinfo{booktitle}{\emph{Proceedings of the Fifth ACM International Conference on Web Search and Data Mining}}.
\newblock


\bibitem[et~al.(2013)]%
        {role_weng_2013}
\bibfield{author}{\bibinfo{person}{Lilian~Weng et al.}} \bibinfo{year}{2013}\natexlab{}.
\newblock \showarticletitle{The role of information diffusion in the evolution of social networks}. In \bibinfo{booktitle}{\emph{Proceedings of Knowledge Discovery and Data Mining}}.
\newblock


\bibitem[et~al.(2010b)]%
        {Cha2010}
\bibfield{author}{\bibinfo{person}{M.~Cha et al.}} \bibinfo{year}{2010}\natexlab{b}.
\newblock \showarticletitle{Measuring user influence in twitter: The million follower fallacy}. In \bibinfo{booktitle}{\emph{Proceedings of the Fourth International AAAI Conference on Weblogs and Social Media}}.
\newblock


\bibitem[et~al.(2010c)]%
        {Kumar2010}
\bibfield{author}{\bibinfo{person}{S.~Kumar et al.}} \bibinfo{year}{2010}\natexlab{c}.
\newblock \showarticletitle{Community detection in large-scale online social networks}. In \bibinfo{booktitle}{\emph{Proceedings of the ACM SIGKDD International Conference on Knowledge Discovery and Data Mining}}.
\newblock


\bibitem[Grabowski and Kosinski(2006)]%
        {evolution_grabowski_2006}
\bibfield{author}{\bibinfo{person}{A. Grabowski} {and} \bibinfo{person}{R. Kosinski}.} \bibinfo{year}{2006}\natexlab{}.
\newblock \showarticletitle{Evolution of a social network: the role of cultural diversity}.
\newblock \bibinfo{journal}{\emph{Physical review. E, Statistical, nonlinear, and soft matter physics}} (\bibinfo{year}{2006}).
\newblock


\bibitem[Hossain et~al\mbox{.}(2024)]%
        {hossain2024evolve}
\bibfield{author}{\bibinfo{person}{Ismail Hossain}, \bibinfo{person}{Md~Jahangir Alam}, \bibinfo{person}{Sai Puppala}, {and} \bibinfo{person}{Sajedul Talukder}.} \bibinfo{year}{2024}\natexlab{}.
\newblock \showarticletitle{EVOLVE: Predicting User Evolution and Network Dynamics in Social Media Using Fine-Tuned GPT-like Model}.
\newblock \bibinfo{journal}{\emph{arXiv preprint arXiv:2407.09691}} (\bibinfo{year}{2024}).
\newblock


\bibitem[Jannach et~al\mbox{.}(2010)]%
        {jannach2010recommender}
\bibfield{author}{\bibinfo{person}{Dietmar Jannach}, \bibinfo{person}{Markus Zanker}, \bibinfo{person}{Alexander Felfernig}, {and} \bibinfo{person}{Gerhard Friedrich}.} \bibinfo{year}{2010}\natexlab{}.
\newblock \bibinfo{booktitle}{\emph{Recommender Systems: An Introduction}}.
\newblock \bibinfo{publisher}{Cambridge University Press}.
\newblock


\bibitem[Jin and Phua(2014)]%
        {Jin2014}
\bibfield{author}{\bibinfo{person}{S.~V. Jin} {and} \bibinfo{person}{J. Phua}.} \bibinfo{year}{2014}\natexlab{}.
\newblock \showarticletitle{Following celebrities’ tweets about brands: The impact of twitter-based electronic word-of-mouth on consumers’ source credibility perception, buying intention, and social identification with celebrities}.
\newblock \bibinfo{journal}{\emph{Journal of Advertising}} \bibinfo{volume}{43}, \bibinfo{number}{2} (\bibinfo{year}{2014}), \bibinfo{pages}{181--195}.
\newblock


\bibitem[Joinson(2008)]%
        {Joinson2008}
\bibfield{author}{\bibinfo{person}{A.~N. Joinson}.} \bibinfo{year}{2008}\natexlab{}.
\newblock \showarticletitle{Looking at, looking up or keeping up with people?: Motives and use of facebook}. In \bibinfo{booktitle}{\emph{Proceedings of the SIGCHI Conference on Human Factors in Computing Systems}}.
\newblock


\bibitem[Lampe et~al\mbox{.}(2006)]%
        {Lampe2006}
\bibfield{author}{\bibinfo{person}{C. Lampe}, \bibinfo{person}{N. Ellison}, {and} \bibinfo{person}{C. Steinfield}.} \bibinfo{year}{2006}\natexlab{}.
\newblock \showarticletitle{A face(book) in the crowd: Social searching vs. social browsing}. In \bibinfo{booktitle}{\emph{Proceedings of the 2006 20th Anniversary Conference on Computer Supported Cooperative Work}}.
\newblock


\bibitem[Li and Chen(2014)]%
        {Li2014}
\bibfield{author}{\bibinfo{person}{Y. Li} {and} \bibinfo{person}{C. Chen}.} \bibinfo{year}{2014}\natexlab{}.
\newblock \showarticletitle{Predicting user behaviors in online social networks}.
\newblock \bibinfo{journal}{\emph{Information Sciences}}  \bibinfo{volume}{317} (\bibinfo{year}{2014}), \bibinfo{pages}{286--298}.
\newblock


\bibitem[Lin et~al\mbox{.}(2009)]%
        {analyzing_lin_2009}
\bibfield{author}{\bibinfo{person}{Yu-Ru Lin}, \bibinfo{person}{Yun Chi}, \bibinfo{person}{Shenghuo Zhu}, \bibinfo{person}{Hari Sundaram}, {and} \bibinfo{person}{Belle~L. Tseng}.} \bibinfo{year}{2009}\natexlab{}.
\newblock \showarticletitle{Analyzing communities and their evolutions in dynamic social networks}.
\newblock \bibinfo{journal}{\emph{ACM Transactions on Knowledge Discovery From Data}} (\bibinfo{year}{2009}).
\newblock


\bibitem[Marwick and Boyd(2011)]%
        {Marwick2011}
\bibfield{author}{\bibinfo{person}{A.~E. Marwick} {and} \bibinfo{person}{D. Boyd}.} \bibinfo{year}{2011}\natexlab{}.
\newblock \showarticletitle{I tweet honestly, I tweet passionately: Twitter users, context collapse, and the imagined audience}.
\newblock \bibinfo{journal}{\emph{New Media \& Society}} \bibinfo{volume}{13}, \bibinfo{number}{1} (\bibinfo{year}{2011}), \bibinfo{pages}{114--133}.
\newblock


\bibitem[McCay-Peet and Quan-Haase(2017)]%
        {McCay-Peet2017}
\bibfield{author}{\bibinfo{person}{L. McCay-Peet} {and} \bibinfo{person}{A. Quan-Haase}.} \bibinfo{year}{2017}\natexlab{}.
\newblock \bibinfo{booktitle}{\emph{What is Social Media and What Questions Can Social Media Research Help Us Answer?}}
\newblock


\bibitem[Ngiam et~al\mbox{.}(2011)]%
        {ngiam2011multimodal}
\bibfield{author}{\bibinfo{person}{Jiquan Ngiam}, \bibinfo{person}{Aditya Khosla}, \bibinfo{person}{Mingyu Kim}, \bibinfo{person}{Juhan Nam}, \bibinfo{person}{Honglak Lee}, {and} \bibinfo{person}{Andrew~Y Ng}.} \bibinfo{year}{2011}\natexlab{}.
\newblock \showarticletitle{Multimodal deep learning}. In \bibinfo{booktitle}{\emph{Proceedings of the 28th International Conference on Machine Learning (ICML-11)}}. \bibinfo{pages}{689--696}.
\newblock


\bibitem[Pariser(2011)]%
        {pariser2011filter}
\bibfield{author}{\bibinfo{person}{Eli Pariser}.} \bibinfo{year}{2011}\natexlab{}.
\newblock \bibinfo{booktitle}{\emph{The Filter Bubble: How the New Personalized Web is Changing What We Read and How We Think}}.
\newblock \bibinfo{publisher}{Penguin}.
\newblock


\bibitem[Radford et~al\mbox{.}(2019)]%
        {radford2019language}
\bibfield{author}{\bibinfo{person}{Alec Radford}, \bibinfo{person}{Jeffrey Wu}, \bibinfo{person}{Rewon Child}, \bibinfo{person}{David Luan}, \bibinfo{person}{Dario Amodei}, {and} \bibinfo{person}{Ilya Sutskever}.} \bibinfo{year}{2019}\natexlab{}.
\newblock \showarticletitle{Language models are unsupervised multitask learners}.
\newblock \bibinfo{journal}{\emph{OpenAI Blog}} \bibinfo{volume}{1}, \bibinfo{number}{8} (\bibinfo{year}{2019}), \bibinfo{pages}{9}.
\newblock


\bibitem[Sharmeen et~al\mbox{.}(2015)]%
        {predicting_sharmeen_2015}
\bibfield{author}{\bibinfo{person}{Fariya Sharmeen}, \bibinfo{person}{TA~Theo Arentze}, {and} \bibinfo{person}{Hjp~Harry Timmermans}.} \bibinfo{year}{2015}\natexlab{}.
\newblock \showarticletitle{Predicting the evolution of social networks with life cycle events}.
\newblock \bibinfo{journal}{\emph{Transportation}} (\bibinfo{year}{2015}).
\newblock


\bibitem[Sun et~al\mbox{.}(2019)]%
        {sun2019videobert}
\bibfield{author}{\bibinfo{person}{Chen Sun}, \bibinfo{person}{Austin Myers}, \bibinfo{person}{Carl Vondrick}, \bibinfo{person}{Kevin Murphy}, {and} \bibinfo{person}{Cordelia Schmid}.} \bibinfo{year}{2019}\natexlab{}.
\newblock \showarticletitle{VideoBERT: A joint model for video and language representation learning}. In \bibinfo{booktitle}{\emph{Proceedings of the IEEE/CVF International Conference on Computer Vision (ICCV)}}. \bibinfo{pages}{7464--7473}.
\newblock


\bibitem[Tsai et~al\mbox{.}(2019)]%
        {tsai2019multimodal}
\bibfield{author}{\bibinfo{person}{Yao-Hung~Hubert Tsai}, \bibinfo{person}{Shaojie Bai}, \bibinfo{person}{Paul~Pu Liang}, \bibinfo{person}{Vijaya~B Kolachalama}, {and} \bibinfo{person}{Louis-Philippe Morency}.} \bibinfo{year}{2019}\natexlab{}.
\newblock \showarticletitle{Multimodal transformer for unaligned multimodal language sequences}. In \bibinfo{booktitle}{\emph{Proceedings of the 57th Annual Meeting of the Association for Computational Linguistics}}. \bibinfo{pages}{6558--6569}.
\newblock


\bibitem[Vaswani et~al\mbox{.}(2017)]%
        {vaswani2017attention}
\bibfield{author}{\bibinfo{person}{Ashish Vaswani}, \bibinfo{person}{Noam Shazeer}, \bibinfo{person}{Niki Parmar}, \bibinfo{person}{Jakob Uszkoreit}, \bibinfo{person}{Llion Jones}, \bibinfo{person}{Aidan~N Gomez}, \bibinfo{person}{{\L}ukasz Kaiser}, {and} \bibinfo{person}{Illia Polosukhin}.} \bibinfo{year}{2017}\natexlab{}.
\newblock \showarticletitle{Attention is all you need}.
\newblock \bibinfo{journal}{\emph{Advances in Neural Information Processing Systems}}  \bibinfo{volume}{30} (\bibinfo{year}{2017}).
\newblock


\end{thebibliography}
\end{document}